\documentstyle[aps,epsf,floats,amsmath,amssymb]{revtex}

\begin{document}
\draft
\renewcommand{\theparagraph}{\Alph{paragraph}}
\twocolumn[\hsize\textwidth\columnwidth\hsize\csname@twocolumnfalse%
\endcsname

\title{Binary spreading process with parity conservation}
\author{Kwangho Park$^{1,3}$, Haye Hinrichsen$^{2}$, and In-mook Kim$^{3}$}

\address{$^{1}$ Theoretische Physik, Fachbereich 10,
     	Gerhard-Mercator-Universit{\"a}t Duisburg,
     	47048 Duisburg, Germany}
\address{$^{2}$ Theoretische Physik, Fachbereich 8, Universit{\"a}t
	GH Wuppertal, 42097 Wuppertal, Germany}
\address{$^{3}$Department of Physics, Korea University, Seoul,
        136-701, Korea}

\date{\today}
\maketitle
\begin{abstract}
Recently there has been a debate concerning
the universal properties of the phase transition
in the pair contact process with diffusion (PCPD)
$2A\to 3A, 2A\to \emptyset$. 
Although some of the critical exponents seem to
coincide with those of the so-called parity-conserving
universality class, it was suggested that the
PCPD might represent an independent class of
phase transitions. This point of view is motivated by
the argument that the PCPD does not conserve parity
of the particle number. 
In the present work we pose the question what happens 
if the parity conservation law is restored. To this end we
consider the the reaction-diffusion process
$2A\to 4A, 2A\to \emptyset$. Surprisingly this process
displays the same type of critical behavior, leading
to the conclusion that the most important characteristics 
of the PCPD is the use of binary reactions for spreading, 
regardless of whether parity is conserved or not.
\end{abstract}
]
%

In the field of nonequilibrium critical phenomena,
the study of phase transitions from fluctuating
into absorbing states continues to attract considerable
attention~\cite{MarroDickman}. It is believed that
phase transitions into absorbing states can be 
categorized into a finite number of universality
classes characterizing the long-range properties
at the critical point.

So far two universality classes are firmly established.
The first and most prominent one is the universality
class of directed percolation (DP)~\cite{Kinzel,Hinrichsen},
which describes the spreading of particles according
to the reaction diffusion scheme
\begin{equation}
A \stackrel{\lambda}{\to} 2A \,, \qquad A 
\stackrel{\mu}{\to} \emptyset \,,
\end{equation}
where $\lambda$ and $\mu$ are the rates for offspring 
production and particle decay, respectively.
In addition, particles are allowed to diffuse
and the maximal density of particles is limited.
Therefore, if $\lambda$ is sufficiently high, the
system is in a fluctuating (active) high-density
phase, while for low values of $\lambda$ it
reaches the (inactive) vacuum state within
exponentially short time.

The second established universality class is the
so-called parity-conserving (PC) class of 
phase transitions~\cite{Tretyakov,CardyTauber,AvrahamRedner} 
which appear in spreading processes with  
parity-conserving dynamics such as
\begin{equation}
A \to 3A \,, \qquad 2A \to \emptyset .
\end{equation}
In this type of spreading process particles can
only annihilate in pairs so that the absorbing phase
is characterized by an {\em algebraic} decay of
the particle density with time. In one spatial   
dimension, parity conservation allows the
particles to be considered as kinks between
oppositely oriented domains~\cite{KimPark,Menyhard94}.
Using this interpretation, the process can be
regarded as a directed percolation process with
two $Z_2$-symmetric absorbing
states (DP2)~\cite{Hinrichsen97}. To some
extent, the situation is similar to the one in the
kinetic Ising model, although in the present
case the transition is generated by
{\em interfacial noise} instead of bulk noise~\cite{Remark}.

Apart from these two established universality
classes, there are only few other possible candidates,
the most mysterious being the pair contact process 
with diffusion (PCPD), sometimes also called
annihilation-fission process. This process 
was originally introduced by Howard and
T\"auber as a model interpolating between 
`real' and `imaginary' noise~\cite{HowardTaebuer}
and corresponds to the reaction-diffusion scheme
\begin{equation}
\label{PCPDReaction}
2A \to 3A \,, \qquad 2A \to \emptyset \,.
\end{equation}
Interestingly, this model exhibits
 a nontrivial phase transition even
in one spatial dimension. As in the PC class,   
particles can only annihilate in pairs, so that
the particle density in the inactive phase decays  
algebraically. Moreover, the model has two   
absorbing states, namely, the empty lattice and
the state with a single diffusing particle.
Because of these similarities and an apparent
numerical coincidence of certain critical exponents,
Carlon {\it et al.} raised the possibility that the
transition in the PCPD might belong to the
PC universality class~\cite{Carlon}. A different point of view 
was presented in Ref.~\cite{Hinrichsen1}, suggesting  that the 
broken parity conservation law in the reaction diffusion
scheme~(\ref{PCPDReaction}) should drive the
system away from the PC class, leading to the
conjecture that the transition in the PCPD might belong to
a novel, yet unexplored universality class. 

Subsequent high-precision simulations~\cite{Odor}
confirmed that some of the critical exponents,
especially the order parameter exponent $\beta$, seem to be
incompatible with the PC hypothesis, supporting
the viewpoint of Ref.~\cite{Hinrichsen1}. On the   
other hand, the simulations revealed unexpected
difficulties, in particular unusually strong corrections
to scaling. It turned out that even after $10^7$
time steps it is not yet clear whether the `true'
scaling regime has already been reached. Therefore,
the critical exponents in one spatial dimension could
only be determined with considerable uncertainty
(see Table~I).

\begin{table}
\begin{tabular}{c||c|c|c}
class & $\beta$ & $\nu_\perp$ & $\nu_\parallel$  \\
\hline
DP 	&0.2765	&1.0969	&1.734 	    \\
PC 	&0.92(2)&1.83(3)& 3.22(6)   \\
\hline
PCPD 	& $< $ 0.6 & 1.0\ldots 1.2	& 1.8 \ldots 2.1  \\
cyclic model 	&0.38(6) & 1.0(1) & 1.8(1)   \\
present work &	0.50(5) &  1.17(7)	& 2.1(1)    	    
\end{tabular}
\vspace{2mm}
\caption{Estimates of the critical exponents for directed percolation,
the parity-conserving class, and various binary spreading processes.}
\end{table}

Concerning the PCPD transition, there are many open
questions: Do the critical properties depend on the
details of the dynamics or do they indeed represent
an independent universality class which has not been
investigated before? Does the simple scaling picture,
which involves only a single length scale, still apply
or is it necessary to consider the possibility of
multiscaling? What is the origin of the scaling
corrections and what are the precise values
of the critical exponents?

A possible phenomenological explanation of the transition
in the PCPD was proposed in~\cite{CyclicPaper}.
This explanation is based on the assumption that
the most salient features of the process are the interplay
of (a) diffusing solitary particles, and (b) spreading   
when at least two particles meet at neighboring
sites. It was conjectured that a cyclically
coupled model with two particle species consisting
of a DP process  and an annihilation process should
display the same critical behavior as the PCPD. 
In fact, numerical estimates of the critical exponents
seem to be compatible with the PCPD results. However,
the mere numerical coincidence within rather large
error bars cannot yet be regarded as a proof.

Since in Ref.~\cite{Hinrichsen1}  the main argument 
against the PC hypothesis has been the broken parity
conservation law in the PCPD, it would be interesting
to find out what happens if the conservation law
is restored. This can be done by modifying the
particle creation process in the
reaction-diffusion scheme~(\ref{PCPDReaction}), 
e.g., by considering the process
\begin{equation}
\label{NewModel}
2A \to 4A\,, \qquad 2A \to \emptyset \,.
\end{equation}
In this process, the number of particles is conserved modulo 2.
As a surprising result, which will be presented below,
we find that this modification does not change the type
of critical behavior at the transition, i.e., the process still
behaves in the same way as the PCPD, as already observed
in the corresponding bosonic field theory~\cite{HowardTaebuer}. 
Thus, in the 
attempt to understand the physics of the PCPD, it would
be misleading to focus exclusively on the parity
conservation law, rather it is more important whether 
we are dealing with a {\em unary} or a {\em binary} spreading process.
In a unary spreading process (e.g. in DP and PC models),
a {\em single} particle is able to produce one or several offspring.
Contrarily, in a binary spreading process such as the PCPD, 
two particles are required to meet at the same or neighboring
sites in order to generate offspring.

\vspace{5mm}
\paragraph{Definition of the model: }
The process defined in~(\ref{NewModel}), which will be studied in the
present work, is a binary spreading process with parity-conserving 
dynamics. It is defined on a one-dimensional lattice with $L$ sites
and periodic boundary conditions, where local variables $s_i= 0,1$
indicate whether the site is empty or occupied by a particle.
The model is controlled by a single 
parameter $p$ and evolves by random-sequential updates 
according to the following dynamic rules. For each update
a site $i$ is randomly selected and a random number $z \in (0,1)$ is
drawn from a flat distribution. Then the following moves are
carried out:
\begin{enumerate}
\item[-] If $p<z$ and site $i$ is occupied by a particle, it hops randomly 
to the left or to the right. If the selected target site is already
occupied, both particles annihilate instantaneously.

\item[-] If $p>z$ and the two sites $i$ and $i+1$ are occupied, this pair of
particles generates two offspring to the left (sites $i-2,i-1$) or
to the right (sites $i+2,i+3$) with equal probability. If the generated
particles land on an already occupied site, they annihilate instantaneously.
\end{enumerate}
\begin{figure}
\epsfxsize=85mm
\centerline{\epsffile{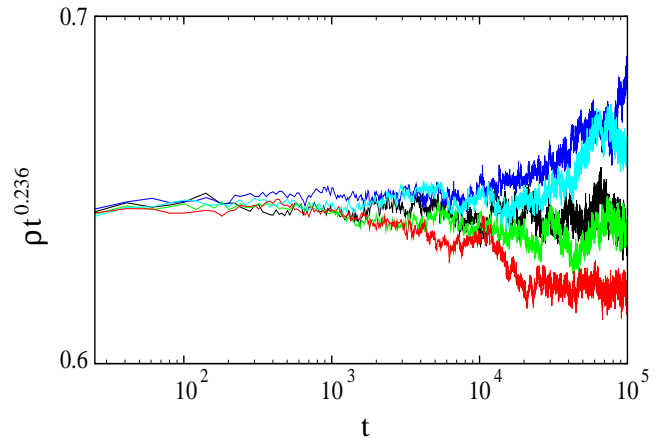}}
\vspace{2mm}
\caption{
\label{FIG1}
The density of particles, $\rho (t)$, times $t^{0.236}$ as a function of time
for $p=0.0893$, $0.0894$, $0.0895$, $0.0896$, and $0.0897$ from top
to bottom, averaged over $2000$ runs on a system with 2048 sizes.
}
\end{figure}
As usual, $L$ update attempts correspond to a time increment of $1$.
The dynamic rules given above
can also be defined in terms of a reaction-diffusion scheme
\begin{equation*}
\begin{array}{cr}
\emptyset A \leftrightarrow A\emptyset & \hspace{3mm} \text{ at rate } p/2 \\
AA \rightarrow \emptyset\emptyset & p \\
AA\emptyset\emptyset \leftrightarrow AAAA & q/2 \\
\emptyset\emptyset AA \leftrightarrow AAAA & q/2 \\
AAA\emptyset \leftrightarrow AA\emptyset A & q/2 \\
\emptyset AAA \leftrightarrow A\emptyset AA & q/2
\end{array}
\end{equation*}
where $q=1-p$.

\vspace{5mm}
\paragraph{Numerical analysis: }

In order to estimate the critical exponents characterizing 
the transition between the active and the absorbing phase
we performed standard Monte Carlo simulations. 
To this end we measured the density of particles, 
$\rho (t) = \frac{1}{L} \sum_i s_i (t)$,
starting with a fully occupied lattice as initial condition.
At the critical point, this quantity should decay algebraically
as $\rho (t) \sim t^{- \delta}$. 
Using this criterion (see Fig.~\ref{FIG1}) we 
estimated the critical point by $p_c =0.0895(2)$.
For the decay exponent we obtain the estimate
\begin{figure}
\epsfxsize=70mm
\centerline{\epsffile{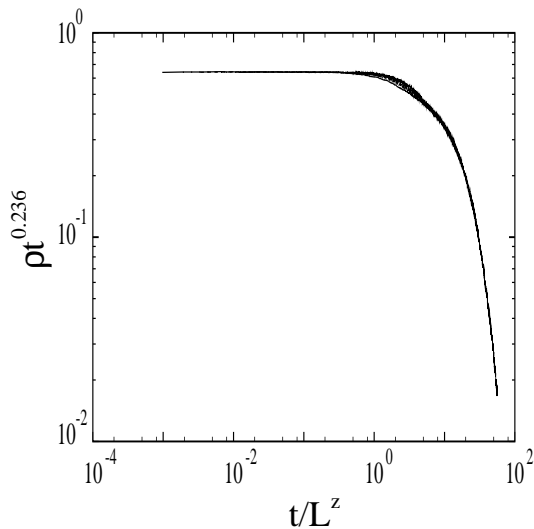}}
\vspace{2mm}
\caption{
\label{FIG2}
Finite size data collapse according to (6) for system sizes
$L=64$, $90$, $128$, $180$, and $256$ averaged over 50000 runs.
}
\end{figure}
\begin{equation}
\delta=\beta/\nu_{\parallel} =0.236(10) .
\end{equation}
Next, in order to determine the dynamic exponent 
$z=\nu_{\parallel}/\nu_{\perp}$, we performed finite size simulations
at the critical point. Here the density of particles should obey the following
finite-size scaling form
\begin{equation}
\rho (t,L) \sim t^{-\delta} f(t/L^z ) ,
\end{equation}
where $f$ is a universal scaling function. 
Using the previous estimate $\delta=0.236$,
the best collapse is obtained
for $ z=1.80(5)$ (see Fig.~\ref{FIG2}). 
Similarly, the third independent exponent 
$\nu_{\parallel}$ can be determined by studying the behavior of the density 
of particles below and above criticality. Here we expect the 
scaling form
\begin{equation}
\rho (t,\epsilon) \sim t^{-\delta} g(t\epsilon^{\nu_{\parallel}} ),
\end{equation}
where $\epsilon = |p - p_c |$ denotes the distance from the critical point.
Using the estimate $\delta=0.236$, the best collapse is obtained 
for $\nu_{\parallel} =2.1(1)$ (see Fig.~\ref{FIG3}).
Combining these estimates we arrive at the result
\begin{equation}
\beta=0.50(5)\,, \qquad \nu_\perp=1.17(7) \,, \qquad \nu_\parallel=2.1(1). 
\end{equation}
As an additional test, we performed
dynamic simulations starting with a seed of a 
single pair of particles located
in the center, measuring the survival 
probability $P(t)$ that the system has not yet reached an absorbing
states, the average number of particles $N(t)$, and the mean
square spreading from the origin $R^2 (t)$ averaged over the survival
runs. At criticality these quantities should obey the power laws
$P(t) \sim t^{-\delta' }$, $N(t) \sim t^\eta$, and $R^2 (t) \sim t^{2/z}$,
where $\delta'$ and $\eta$ are dynamical exponents. 
Measuring these quantities we observe strong corrections to scaling, which
make it impossible to obtain reliable estimates for the critical exponents.
Fitting straight lines over the last decade we find the values
(see Fig.~\ref{FIG4})
\begin{equation}
\delta' \approx  0.1, ~~~ \eta \approx 0.2, ~~~ 2/z  \approx 1.15 
\end{equation}
without being able to estimate the errors.  Nevertheless the estimate for 
$2/z$ is in rough agreement with the previous estimate $z=1.80(2)$. 

We also measured the density of {\it pairs} of particles, which can be
used as an alternative order parameter in the present model. Here we find
the same type of critical behavior, although with slightly different estimates
for the critical exponents.
\begin{figure}
\epsfxsize=70mm
\centerline{\epsffile{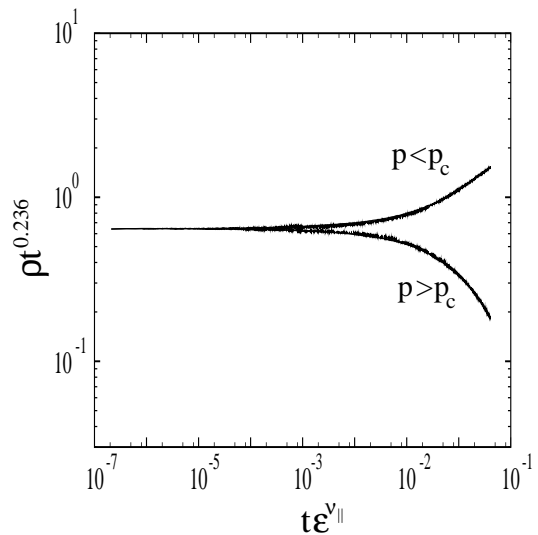}}
\vspace{2mm}
\caption{
\label{FIG3}
Data collapse for off-critical simulations according
to the scaling form (7) for $\epsilon = 0.0001$, $0.0002$,
$\ldots$, $0.0064$ averaged over 2000
runs.                                      
}
\end{figure}
%
%
%
\vspace{5mm}
\paragraph{Discussion: }

As shown in Table 1, our estimates for the critical exponents 
are in fair agreement with those of the PCPD. In particular, we can rule 
out the  possibility of a DP or a PC transition. 
This result is surprising since it suggest that we can introduce 
an additional symmetry without changing the critical behavior 
of the transition. This means that parity conservation is {\em irrelevant} 
for the long-range properties at the transition.

To understand this observation, we note that there is another
well-known example where parity conservation is irrelevant,
namely, the annihilation process $2A \rightarrow 0$  in comparison to
the coagulation process $2A \rightarrow A$. 
Both processes are known to belong to the same universality class and can even
be related by an exact similarity transformation~\cite{EquivalenceAnnhCoag}. 
This is due to the fact 
that the even and the odd sector in the parity-conserving
process $2A \rightarrow 0$ are essentially equivalent since
in both cases the particle density decays algebraically until 
the system is trapped in an absorbing state (namely, the empty lattice 
or a state with a single diffusing particle). Breaking the
parity conservation law by a weak perturbation 
the system begins to switch between
the even and the odd sector. However, this ongoing switching process
does not change the universal behavior 
since -- from a macroscopic point of view --
the physical properties of both sectors cannot be distinguished.
\begin{figure}
\epsfxsize=85mm
\centerline{\epsffile{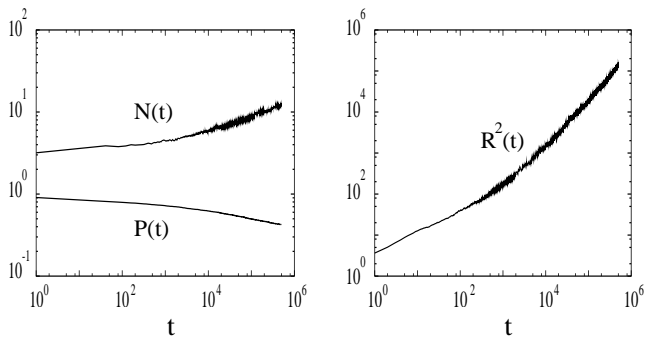}}
\vspace{2mm}
\caption{
\label{FIG4}
The survival probability $P(t)$, the average number of particles
N(t), and the mean square spreading $R^2 (t)$ starting with a single
pair of particles.
}
\end{figure}
In the present model the situation is quite similar. In both
sectors we have a transition from an active phase into an
absorbing state. Therefore, the physical properties of both 
sectors are essentially the same so that the breakdown of
parity conservation does not change the critical behavior.

In the PC class, however, the situation is completely different.
In this case parity conservation is indeed relevant. For example,
in the branching-annihilating random walk with even number
of offspring $A \to 3A$, $2A\to\emptyset$ the two sectors
are not equivalent because only one of them has an absorbing state.
Therefore, even a tiny violation of the conservation law drives
the transition away from the PC class.

How can we verify whether parity conservation in a given system
is relevant or not? One way would be to investigate how the
critical behavior changes if the symmetry is broken. Another
much more elegant method would be to 
compare seed simulations in the even and
the odd sector (i.e., starting with two or three particles). Here
the survival probability $P(t)$ has to be defined as the probability 
that there are at least two particles left. If the survival exponent
$\delta'$ and the critical initial slip exponent $\eta$ are different
in both sectors (as they are in the case of the PC class), 
parity conservation is relevant. However, if the exponents do
not depend on the sector (as in the present model), we expect
the parity symmetry to be irrelevant.

In the light of these results, the conjecture of Ref.~\cite{Hinrichsen1}
has to be refined. It is true that we cannot have PC critical behavior
in systems without parity conservation or an equivalent $Z_2$ symmetry.
On the other hand, the broken parity conservation law is not the main 
characteristic of the transition in PCPD, rather it is possible to restore
this symmetry without changing the critical behavior. Therefore,
a necessary condition for existence of this class, for which our understanding is
still incomplete, seems to be the {\em binary} nature of the
process for offspring production, i.e., 
two particles have to meet at the same place
in order to create new particles.

\vspace{3mm}
This work is supported in part by postdoctoral fellowship program
and Grant No. 98-702-05-01-3 from the Korean Science and Engineering 
Foundation(KOSEF), and also in part by the Ministry of Education through 
the BK21 project.


\begin{references}
\bibitem{Remark} This universality class is sometimes referred to as the
 Directed Ising class.
\end{references}
\end{document}